\documentclass[aps,prc,groupedaddress,showpacs,twocolumn,floatfix]{revtex4}

\usepackage{graphicx}

\def\gtorder{\mathrel{\raise.3ex\hbox{$>$}\mkern-14mu
 \lower0.6ex\hbox{$\sim$}}}
\def\ltorder{\mathrel{\raise.3ex\hbox{$<$}\mkern-14mu
 \lower0.6ex\hbox{$\sim$}}}

\def\gegm{G_E / G_M}
\def\ge{G_E}
\def\gm{G_M}
\def\etal{\textit{et al.}}

\begin{document}

\title{Evidence for two-photon exchange contributions in
electron-proton and positron-proton elastic scattering}

\author{J. Arrington}

\affiliation{Physics Division, Argonne National Laboratory, Argonne, Illinois 60439, USA}

\date{\today}

\begin{abstract}

The comparison of positron-proton and electron-proton elastic scattering cross
sections is a sensitive test for the presence of two-photon exchange
contributions. Thirty years ago, positron data were considered adequate to set
tight limits on the size of two-photon corrections.  More recently, these
radiative corrections have again become a matter of great interest as a
possible explanation for the discrepancy between Rosenbluth and polarization
transfer measurements of the proton electromagnetic form factors.  We have
reexamined the electron and positron scattering data to see if they can
accommodate two-photon effects of the size necessary to account for the
Rosenbluth-polarization transfer discrepancy. The data are consistent
with simple estimates of the two-photon contributions necessary to explain the
discrepancy. In fact, they strongly favor a large $\varepsilon$-dependent correction to
the positron to electron ratio, providing the first direct experimental
evidence for a two-photon contribution to unpolarized lepton-proton scattering.

\end{abstract}

\maketitle


\section{INTRODUCTION}\label{sec:intro}

Measurements of $\gegm$, the ratio of the proton electric and magnetic form
factors, from Rosenbluth separation and polarization transfer techniques yield
significantly different results~\cite{arrington03} at large values of $Q^2$,
the four-momentum transfer squared. The systematic uncertainties of both the
Rosenbluth~\cite{arrington03} and polarization~\cite{punjabi03} measurements
have been studied in detail, and no explanation for the discrepancy in terms
of experimental problems has been found. If the discrepancy is not due to
errors in the experiments or analyses, it may indicate a more fundamental
problem with one of the techniques.  Until this discrepancy is understood,
there will be large uncertainties in our knowledge of the proton form factors.
Since the polarization transfer measurements have only extracted the ratio
$\gegm$, cross section measurements are still needed to determine the
magnitude of the individual form factors.  So even if it is shown that the
polarization transfer measurements are correct, and that the problem is due to
unaccounted for corrections in the cross section measurements, there will
still be uncertainties in the form factors until we fully understand these
corrections~\cite{arrington03b}.

Because the discrepancy grows rapidly with $Q^2$, it has typically been
assumed that it is a problem with the cross section measurements, where a
fixed error in the $\varepsilon$ dependence of the cross sections would yield
an error in $(G_E/G_M)^2$ that grows approximately linearly with $Q^2$.
Assuming that the difference is due primarily to missing corrections in the
cross section measurements, the discrepancy requires an error in the
$\varepsilon$ dependence of the cross section of approximately 5--8\% for $1 <
Q^2 < 6$~GeV$^2$~\cite{arrington03, arrington03b, guichon03}.

In order for a modification to the cross section to change the extracted form
factor ratio, it must modify the $\varepsilon$ dependence of the reduced cross
section,
\begin{equation}
\label{eq:rosenbluth}
\sigma_R \equiv
\frac{d\sigma}{d\Omega} \frac{\varepsilon(1+\tau)}{\sigma_{\rm Mott}}
= \tau \gm^2(Q^2) + \varepsilon \ge^2(Q^2),
\end{equation}
where $\tau=Q^2/4 M_p^2$, and $\varepsilon$ is the longitudinal polarization
of the virtual photon [$\varepsilon^{-1} = 1+2(1+\tau)\tan^2{(\theta/2)}$].
Several attempts have been made to find effects that might introduce an
additional $\varepsilon$ dependence to the measured cross section, thus
modifying the extracted Rosenbluth form factors. Coulomb corrections, when
implemented in a simple effective momentum approximation~\cite{higinbotham03},
do modify the $\varepsilon$ dependence of the cross section, but yield a very
small effect compared to the size needed to explain the discrepancy.  For the
most part, investigations have focussed on the effect of two-photon exchange
corrections~\cite{guichon03, blunden03, rekalo03} beyond the limited
contributions that are already included in the traditional calculations of
radiative corrections~\cite{mo69, tsai71, meister63}.

While these works have shown that it is possible that a two-photon correction
could explain the discrepancy, the only quantitative
calculation~\cite{blunden03} is limited to the elastic part of the two-photon
contributions, i.e., the box and crossed-box diagrams considering only
the case where the intermediate state is a proton, and yield only a 2\%
$\varepsilon$ dependence, less than half the size necessary to explain the
discrepancy.  In this work, we reexamine positron measurements that were
designed to test for two-photon contributions in elastic scattering in light
of the possibility that they may be responsible for the discrepancy.

\section{EXPERIMENTAL LIMITS ON TWO-PHOTON CONTRIBUTIONS}

The effect of two-photon exchange terms can be observed in several
processes.  The imaginary part of the two-photon amplitude
can, in principle, be measured in polarization observables.  Measurements
of the normal polarization, $P_N$, which is zero in the Born approximation,
have been made~\cite{bizot65, giorgio65, lundquist68}, but no statistically
significant indication of two-photon contributions has been seen. Similarly,
the asymmetry $A_N$ has been measured for both elastic and inelastic
scattering~\cite{chen68, powell70}, again with only null results.  Thus far,
the only observations of possible two-photon effects are in the asymmetry
of scattering of transversely polarized electrons from protons. These
asymmetries are extremely small, of order $10^{-5}$, and so extremely
difficult to measure.  However, they have been observed by the
SAMPLE experiment at MIT-Bates~\cite{wells01} and the PVA4 Collaboration at
Mainz~\cite{maas_priv}.  However, these polarization observables
are related to the imaginary part of the two-photon amplitude, while the
cross section measurements are related to the real part.  Therefore, while
these data can be used to test models of the two-photon exchange, they do 
not directly constrain the two-photon contributions to the unpolarized cross
sections, which might explain the discrepancy.

There are two ways to look for the effects of two-photon exchange corrections
in the unpolarized elastic electron-proton cross section.  First, one can look
for deviations from the linear $\varepsilon$ dependence in
Eq.~\ref{eq:rosenbluth}. There are a few Rosenbluth separation
measurements~\cite{borkowski74, andivahis94, christy03} that cover both large
and small $\varepsilon$ values and have small uncertainties (1--2\%).
These measurements do not show any significant deviations from linearity,
but they have limited sensitivity because they have little data below
$\varepsilon=0.4$ and no data below $\varepsilon=0.2$. Data from different
experiments can be combined to expand the $\varepsilon$ range, but
normalization uncertainties between different experiments reduce the
significance of such tests, while attempts to normalize across the
datasets~\cite{walker94,arrington03} rely on the linearity when determining
normalization factors.  So while these data set limits on non-linearities at
large $\varepsilon$, the limits are less significant at low $\varepsilon$.  In
addition, these measurements are insensitive to two-photon correction terms
that are constant or vary linearly with $\varepsilon$.  Such corrections
would modify the extracted values of $\ge$ and $\gm$ without spoiling the
linearity of the Rosenbluth plot.

The second approach is to compare positron-proton scattering to
electron-proton scattering.  For positron-proton scattering, the interference
term between the one-photon and two-photon amplitudes changes sign, yielding
a ratio $R \equiv \sigma(e^+p)/\sigma(e^-p) \approx 1 + 4 \mbox{Re}(B)/A$,
where $B$ is the two-photon amplitude, and $A$ is the one-photon
amplitude~\cite{mar68}. The modification to the electron cross section is
approximately $1 - 2\mbox{Re}(B)/A$, and so any change in the electron cross
section will yield roughly twice the change in $R$, but with the opposite sign.
In the simplest approximation, one expects an additional factor of $\alpha$ in
the two-photon amplitude relative to one-photon amplitude, yielding
corrections to the electron cross section of roughly $2\alpha \approx 1.5\%$,
and to the ratio R of $\approx 3$\%. Additional differences come from
Bremsstrahlung corrections where proton recoil is taken into account, but
these are included in the usual radiative corrections.  An analysis by Mar and
collaborators~\cite{mar68} found these additional differences to be relatively
small, typically less than 1--2\% for their kinematics, and to be identical to
better than 0.3\% in different prescriptions~\cite{meister63, tsai61} of the
radiative corrections.

\begin{figure}[thb]
\includegraphics[height=5.0cm,angle=0]{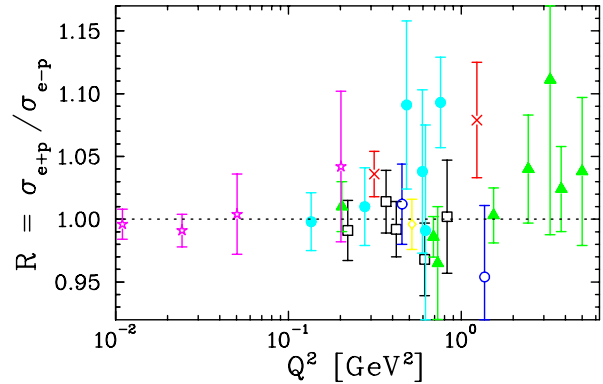}
\caption{(Color online) The cross section ratio, $R = \sigma_{e+} /
\sigma_{e-}$,  as a function of $Q^2$.  The experiments are differentiated by
color and symbol: black squares~\cite{anderson68}, red
crosses~\cite{bouquet68}, green solid triangles~\cite{mar68}, blue hollow
circles~\cite{bartel67}, yellow diamonds~\cite{anderson66}, cyan filled
circles~\cite{browman65}, and magenta stars~\cite{yount62}.}
\label{fig:positron_qsq}
\end{figure}

Figure~\ref{fig:positron_qsq} shows the existing data~\cite{yount62,
browman65, anderson66, bartel67, cassiday67, anderson68, bouquet68, mar68} for
the ratio of positron-proton to electron-proton elastic cross sections as a
function of $Q^2$. While there is some hint of a $Q^2$ dependence, the large
$Q^2$ data have large uncertainties, and a fit to the data of the form $R = a +
b Q^2$ yields $b=0.0085\pm0.0063$, less then 1.5 standard deviations from
zero. A fit of the ratios to a constant value yields $\langle R \rangle =
1.003 \pm 0.005$, with $\chi^2_\nu = 0.87$, which corresponds to a two-photon
correction to the electron cross sections of ($-0.15\pm0.25$)\%. This result
has been interpreted to mean that the two-photon corrections must be even
smaller than the naive estimate, limiting the effect on the electron-proton
cross section to less than one percent.

However, the low intensity of the secondary positron beams used in these
experiments makes it difficult to perform precise measurements where the cross
section is small. Because of this, the data in Fig.~\ref{fig:positron_qsq} are
limited to low $Q^2$ values ($\ltorder 1$~GeV$^2$) or small scattering angles
($\varepsilon > 0.7$), where the cross section is large.  While the existing
data do place tight limits on the size of two-photon corrections in some
regions, they do not place any limits on two-photon contributions at low
$\varepsilon$ except at relatively low $Q^2$ values ($\ltorder$1 GeV$^2$).  So
it is still possible that the discrepancy in the extracted form factors is due
to two-photon corrections to the cross sections, if the correction is only
large for small $\varepsilon$ values.

If we assume that the two-photon corrections are responsible for the
discrepancy between polarization transfer and Rosenbluth measurements, we can
make specific predictions about how these corrections would impact the
positron measurements, and use this to examine the existing data more
carefully. In order to explain the discrepancy, the effect must increase the
slope of the Rosenbluth plot, and so must increase the cross section at large
$\varepsilon$ relative to the low $\varepsilon$.  Based on the size and
$Q^2$ dependence of the discrepancy, the $\varepsilon$ dependence in the
electron cross section must be 5--8\%, depending only weakly on $Q^2$, for
$Q^2 \gtorder 2$~GeV$^2$. It must also be reasonably close to linear in
$\varepsilon$, or else it would introduce visible non-linearities in the
Rosenbluth plot.  This implies that the ratio $R$ should have a 10--15\%
$\varepsilon$ dependence, approximately linear in $\varepsilon$, and of the
opposite sign as in the electron cross section, i.e., the positron to
electron ratio must either increase at small $\varepsilon$ or decrease at
large $\varepsilon$.

Unfortunately, there is very little positron data above $Q^2$=2,
and it covers a very limited $\varepsilon$ range.  The data from Mar,
\etal~\cite{mar68}, has four points above $Q^2=2$~GeV$^2$, yielding $\langle R
\rangle = 1.034 \pm 0.024$. These data are all at large $\varepsilon$ values
($\langle \varepsilon \rangle = 0.88$), and so do not exclude
significant two-photon corrections at large $Q^2$. If the two-photon
correction is small at $\varepsilon=0$, then a 10--15\% decrease in $R$ at
large $\varepsilon$ would be necessary to explain the discrepancy, and this is
clearly ruled out by the high $\varepsilon$ data. So any two-photon
corrections would have to increase $R$ (decrease the electron cross section)
at low $\varepsilon$ in order to explain the discrepancy and still be
consistent with the positron data.

\begin{figure}[thb]
\includegraphics[height=5.0cm,angle=0]{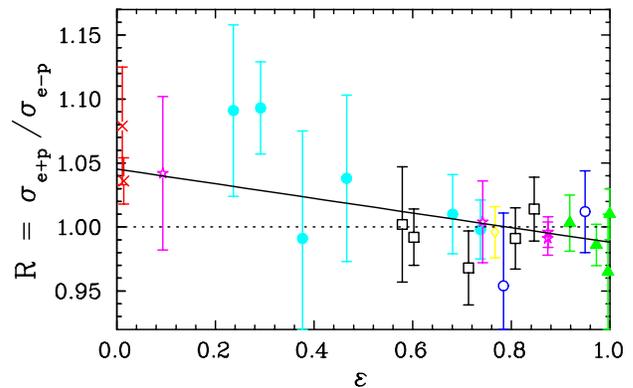}
\caption{(Color online) $\sigma_{e+} / \sigma_{e-}$ cross section ratio as a
function of $\varepsilon$ for the measurements below $Q^2=2$~GeV$^2$.  The
solid line is a fit assuming a linear $\varepsilon$ dependence and no
$Q^2$ dependence to the ratio, and yields a slope of $-(5.7 \pm 1.8)$\%. The
symbols are identical to Fig.~\ref{fig:positron_qsq}.}
\label{fig:positron_eps}
\end{figure}

The positron data with significant $\varepsilon$ range is limited to $Q^2 <
2$~GeV$^2$. Figure~\ref{fig:positron_eps} shows these data as a function of
$\varepsilon$, and a significant $\varepsilon$ dependence can be seen. 
A linear fit, neglecting any $Q^2$ dependence,
yields an slope of $-(5.7 \pm 1.8)$\%, with $\chi^2=11.1$ for 22 degrees of
freedom.  The extremely low $\chi^2$ indicates that the uncertainties in the
data have most likely been overestimated, and that the effect may be more
significant than indicated by the fit uncertainty.

The observed increase in the positron to electron ratio at low $\varepsilon$
corresponds to an increase of 2.8\% in the observed slope in the Rosenbluth
extraction. This implies that the value of $G_E$ extracted from the Rosenbluth
separation will be larger than the true value, while the extracted $G_M$ value
will be smaller. While the 2.8\% $\varepsilon$ dependence is only half the
size necessary to explain the discrepancy at large $Q^2$ value, data covering
a wide range of $\varepsilon$ values are only available at low $Q^2$. The
average $Q^2$ value of the data in Fig.~\ref{fig:positron_eps} is only
0.5~GeV$^2$, and the lower $Q^2$ data are generally more precise, making the
weighted average $Q^2$ value less than 0.4~GeV$^2$.

We can estimate the $\varepsilon$ dependence necessary to explain the
discrepancy in the form factors at large $Q^2$, but at these low $Q^2$ values
the polarization transfer and Rosenbluth form factors are not precise enough
to determine if there is an inconsistency, and so cannot be used to estimate
the size of the two-photon corrections. A decrease in the size of the
$\varepsilon$ dependence at low $Q^2$ could easily yield the slope observed in
Fig.~\ref{fig:positron_eps}, yet still be large enough to fully explain the
discrepancy between polarization and Rosenbluth extractions at larger $Q^2$
values. At larger $Q^2$ values, where the size of the corrections can be
estimated from the discrepancy, the effect decreases somewhat as $Q^2$
decreases, and is approximately 5\% for $Q^2 =1$--2~GeV$^2$.  In addition, the
correction must become smaller for very small $Q^2$ values
($0.01-0.1$~GeV$^2$), or the decrease in the low-$\varepsilon$ cross sections
would lead to significant reductions in the extracted values of $\gm$.  The
extractions of $\gm$ are not precise enough to conclude that the corrections
must go to zero, but they must be significantly smaller than the 5\%
corrections observed at larger $Q^2$ values. Thus, we expect this slope in the
positron to electron comparison to be less than 10\%.

\begin{figure}[thb]
\includegraphics[height=5.0cm,angle=0]{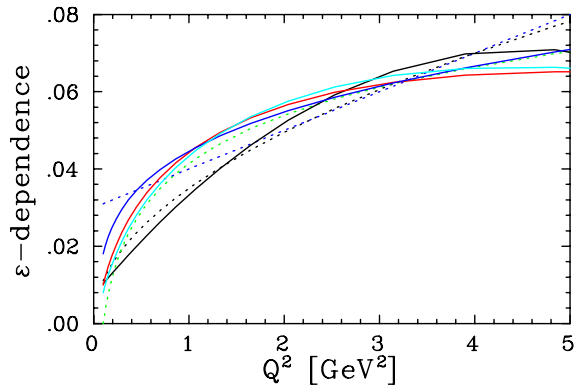}
\caption{(Color online) The $\varepsilon$ dependence of the electron-proton
cross section as a function of $Q^2$ estimated from the discrepancy between
cross section and polarization transfer measurements.  The four curves
correspond to four different parametrizations for the $Q^2$ dependence.
The $\varepsilon$ dependence of the positron-to-electron ratio
should be of opposite sign and approximately twice the size of the
$\varepsilon$ dependence in the electron cross section.}
\label{fig:twophotonfits}
\end{figure}

A global analysis of the cross section and polarization transfer data was used
to try and estimate the low $Q^2$ behavior.  In Ref.~\cite{arrington03b}, a
global analysis of the cross section and polarization transfer data, assuming
a fixed 6\% $\varepsilon$-dependent correction to the cross section, was used
to extract the `Polarization form factors'.  A modified version of this global
analysis was performed, but rather than extracting $\ge$ and $\gm$ with a fixed
two-photon correction, we extract $\ge$, $\gm$, and the $Q^2$ dependence of
the slope of the linear $\varepsilon$-corrections.  Several different
functional forms were tried, and a range of curves, which all gave good fits,
are shown in Fig.~\ref{fig:twophotonfits}.  While the fits were not constrained
to go to zero, they all yield a much smaller value as $Q^2 \rightarrow 0$.
When these curves are used to estimate the $\varepsilon$ dependence for the
correction to the electron cross section at $Q^2=0.4$~GeV$^2$, they yield
slopes of (1.8--3.3)\%, implying a slope in the positron to electron ratio of
$-$(3.7--6.8)\%, in agreement with the observed $-5.7$\% slope.

The $\varepsilon$ dependence extracted above assumed no $Q^2$ dependence to
the size of the correction, and a simple linear $\varepsilon$ dependence. 
While it is in agreement with the estimated $\varepsilon$ dependence from the
form factor discrepancy, the estimate relied on an extrapolation to lower
$Q^2$ values. However, while the above analysis had to make some assumptions
about the $Q^2$ dependence of the two-photon effects, we can also make some
significant model-independent statements from this data.

For the low $\varepsilon$ data ($\varepsilon < 0.5$) $\langle Q^2 \rangle = 0.5$~GeV$^2$ (weighted
average) and $\langle R \rangle = 1.049 \pm 0.014$, clearly demonstrating
that two-photon effects decrease the electron cross section at low
$\varepsilon$ and low $Q^2$. Note that only one point below $\varepsilon=0.5$
is above $Q^2=1$~GeV$^2$, and it has a positron to electron ratio of $1.079
\pm 0.046$.  This is consistent with the 10--12\% increase that would explain
the discrepancy in form factor measurements for $Q^2 = $1--2~GeV$^2$,
but also only two standard deviations from $R=1$.

In addition to the observation of a significant $\varepsilon$ dependence at
low $Q^2$ values, the data also sets significant limits on possible two-photon
exchange corrections at large $Q^2$ for $\varepsilon \gtorder 0.8$. For $Q^2 >
1$~GeV$^2$, $\langle R \rangle=1.020\pm.015$.  So the 95\% confidence region
for $R$ at large $\varepsilon$ is 0.99--1.05, yields limits on the electron
cross section modification of -2.5\% to +0.5\%.  To increase the slope of the
Rosenbluth plot, they would have to increase the high-$\varepsilon$ electron
cross section, and such an enhancement is limited to $<0.5$\%.

\section{IMPLICATIONS OF THE POSITRON MEASUREMENTS}

The $\varepsilon$ dependence of the two-photon effects seen here is consistent
with the calculations of Refs.~\cite{blunden03,afanasev_priv}, which have small
corrections at large $\varepsilon$, and a significant decrease of the electron
cross section at low $\varepsilon$. However, it rules out the form of
Ref.~\cite{rekalo03} as an explanation for the discrepancy between
polarization and Rosenbluth extractions of $\gegm$.  In Ref.~\cite{rekalo03},
the authors predict a specific $\varepsilon$ dependence, which is zero at
$\varepsilon = 0$, and grows rapidly at large $\varepsilon$.  If the size of
the correction is made small enough to be consistent with the constraints from
the positron measurements at large $\varepsilon$, then the two-photon effects
at smaller $\varepsilon$ values will be negligible.

The positron measurements are also inconsistent with the corrections obtained
in Ref.~\cite{guichon03}, at least given the specific approximations that the
authors use to obtain the two-photon effects from the discrepancy in form
factor measurements.  They do not assume single photon exchange, but instead
write a more general expression for the cross section in terms of two
generalized form factors, $\widetilde{G}_E$ and $\widetilde{G}_M$, along with
a third term, $\widetilde{F}_3$, which is zero in the Born approximation. By
assuming that the two-photon contributions are negligible in $\widetilde{G}_E$
and $\widetilde{G}_M$, they extract values of $Y_{2\gamma}$, a dimensionless
parameter related to the size of $\widetilde{F}_3$, such that the effect of
$Y_{2\gamma}$ on the cross section and polarization transfer data resolve the
discrepancy. Under this assumption, the two-photon effects on the cross
section measurements are approximately proportional to $\varepsilon$, and are
$\gtorder$5\% at $\varepsilon=1$ for $Q^2>1$~GeV$^2$. This would yield $R
\approx 0.9$ for the large $Q^2$ positron measurements, which is clearly ruled
out by the data (Fig.~\ref{fig:positron_qsq}).  It is possible that the
two-photon effects in their $Y_{2\gamma}$ terms could be cancelled by
two-photon effects that modify $\widetilde{G}_E$ and $\widetilde{G}_M$.  If
this is the case, the formalism may still allow a connection between the
two-photon effects in polarization transfer and cross section measurements,
but it is no longer possible to determine the two-photon terms directly from
the extraction of $Y_{2\gamma}$, because of the sizable two-photon
contributions to $\widetilde{G}_E$ and $\widetilde{G}_M$.

This observation of the form of the two-photon effects can also be used to
assist in the extraction of form factors from Rosenbluth and polarization
transfer data.  To have a consistent extraction of the form factors from the
cross section and polarization data, we have to assume something about the
nature of the discrepancy.  An $\varepsilon$ dependence of the cross section,
of the form observed in the positron data, is consistent with the assumption
used in Ref.~\cite{arrington03b}. In this case, it was assumed that the cross
sections were modified by two-photon exchange terms that were zero at
$\varepsilon=1$, linear in $\varepsilon$, and large enough to explain the
discrepancy. This assumed modification to the cross sections was removed to
correct for the two-photon effects, with no correction at $\varepsilon=1$, and
a 6\% increase in the cross sections at small $\varepsilon$ ($\sigma_{c2}$ of
Ref.~\cite{arrington03b}). The size of this correction is such that the
Rosenbluth data approximately reproduce the polarization transfer values of
$\gegm$, and the 6\% increase in the $\varepsilon=0$ cross section yields a
value of $\gm$ that is approximately 3\% higher than a direct Rosenbluth
extraction from the unmodified cross sections (e.g. the
parametrization of Ref.~\cite{bosted94} or the `Rosenbluth form factors' of
Ref.~\cite{arrington03b}).

A similar combined extraction of form factors in Ref.~\cite{brash02} used the
polarization transfer values for $\gegm$ to fix the slope of the reduced
cross section, and used the uncorrected cross sections to extract the
magnitude of $\ge$ and $\gm$. For a data point at $\varepsilon=1$, the change
in the assumed slope of Eq.~\ref{eq:rosenbluth} yields an increase in the
extrapolation to $\varepsilon=1$ of 5--8\%, the size implied by the
discrepancy, and so gives similar results to the extraction of
Ref.~\cite{arrington03b}.  For a measurement at very low $\varepsilon$, the
change in slope leaves the $\varepsilon=0$ extrapolation unchanged.  So
depending on the mean $\varepsilon$ value of the data in a given $Q^2$ range,
the extracted value of $\gm$ will be 0--4\% lower in this analysis, compared
to Ref.~\cite{arrington03b}, with a typical difference of 1--2\%, as there are
more data at large $\varepsilon$ values. Note that both of these combined
extractions rely on the assumption that the two-photon exchange terms have
little or no effect on the polarization transfer results.

\section{CONCLUSIONS}

This is the first direct experimental evidence for large two-photon
corrections in the unpolarized elastic electron-proton cross section.  The
effect is only observed for low $Q^2$ values, and so cannot be directly
compared to the two-photon contributions necessary to explain the discrepancy
between Rosenbluth and polarization transfer measurements of the proton form
factors. However, the size and $\varepsilon$ dependence of these effects are
consistent with simple estimates based on the observed discrepancy, and so this
observation supports the idea that two-photon contributions may significantly
modify the Rosenbluth extraction of nucleon form factors.

Additional comparisons of positron to electron scattering over a range in
$Q^2$ and $\varepsilon$ would provide the most direct extraction of these
two-photon corrections.  With precise measurements over an adequate range in
$\varepsilon$ and $Q^2$, we could determine if the two-photon effects can
fully explain the difference between polarization transfer and Rosenbluth
measurements of the form factors, and could also provide significant data with
which to constrain models of the real part of the two-photon amplitude.
However, at the present time it is unclear where such a program could be
carried out over the necessary kinematic range, and with the precision needed
to map out these corrections.  In the meantime, the existing data can be used
to test calculations of the two-photon effects, and already are sufficient to
rule out approaches with large electron cross section enhancements at large
$\varepsilon$ values.  These data also provide information about the size and
$\varepsilon$ dependence of the two-photon effects, important when attempting
to extract the proton form factors from a combined analysis of Rosenbluth and
polarization transfer data.

\begin{acknowledgments}

This work was supported by the U. S. Department of Energy, Nuclear Physics
Division, under contract W-31-109-ENG-38.

\end{acknowledgments}

\bibliography{positron}

\end{document}